\documentclass[aps,twocolumn,pra,showpacs,amsmath,amssymb,showkeys,10pt]{revtex4-2}
\usepackage{graphicx}
\usepackage{epstopdf}
\usepackage{diagbox}
\pdfoutput=1
\usepackage{braket}
\usepackage{hyperref}
\usepackage{longtable}
\usepackage{float}
\usepackage{color}

\begin{document}

	\title{Studying the effect of phonon coherence and inflow on hydrogen bond formation in the $[(\mathrm{H}_2\mathrm{O})_2]^m$ cluster}
	
	\author{Hui-hui Miao}
	\email[Correspondence to: Vorobyovy Gory 1, Moscow, 119991, Russia. Email address: ]{hhmiao@cs.msu.ru}
	\affiliation{Faculty of Computational Mathematics and Cybernetics, Lomonosov Moscow State University, Vorobyovy Gory 1, Moscow, 119991, Russia}

	\date{\today}

	\begin{abstract}
	We propose a simplified open-quantum-system model for hydrogen-bond formation in water clusters, where each subsystem is mapped to a $\lambda$-type three-level system coupled to two effective phonon modes: a micro-vibration mode ($\Omega_{\mathrm{hyd}}$) representing the O--H stretching vibration, and a macro-displacement mode ($\Omega_{\mathrm{dist}}$) representing the intermolecular donor--acceptor motion. The $[(\mathrm{H}_2\mathrm{O})_2]^m$ cluster is studied for $m=2$ to $6$ in the incoherent case (independent phonon modes) and the coherent case (shared phonon modes). We find that phonon coherence significantly alters the dynamics. In the dissipative case, coherence induces a redistribution of steady-state populations: intermediate hydrogen-bond counts are enhanced while edge counts are suppressed --- a ``squeezing'' effect explained by the interplay of subradiant states and dark states. For $m\ge 3$, true dark states emerge, rooted in the permutation symmetry of the system. Inflow of $\Omega_{\mathrm{dist}}$ phonons promotes hydrogen bond formation, while inflow of $\Omega_{\mathrm{hyd}}$ phonons inhibits it. Our results reveal a nontrivial role of quantum coherence and dark states in hydrogen-bond dynamics, providing a foundation for extending the framework to more complex systems.
	\end{abstract}

	

	\maketitle

	\section{Introduction}
	\label{sec:Intro}
	
	One of the most crucial areas in polymer chemistry and macromolecular biology is the modeling of complex quantum systems~\cite{McArdle2020, Albuquerque2021, Baiardi2023}. In this field, the creation and breaking of hydrogen bonds---discovered in Ref.~\cite{Moore1912} and first described in Ref.~\cite{Latimer1920}---is the most significant kind of chemical transformation. Proton tunneling between two conventional potential wells of distinct molecules generates these bonds. External variables can significantly influence the formation of hydrogen bonds due to the relative weakness of these bonds compared to covalent bonds. The complexity issue is made worse by the fact that, unlike the molecular dynamics of ready-made molecules, the dynamics of chemical transformations necessitates the presence of an electromagnetic field. Recently, there has been a surging interest in hydrogen bonds in chemical and biological systems, for example, in decoherence~\cite{Ignacio2023}, entangled spin states~\cite{He2022}, the $\alpha$-helix of proteins~\cite{Danko2022}, liquids and water clusters~\cite{Mei1998, Ceotto2017, Di2018, Yamada2020, You2025}, and enzymes~\cite{Farrow2018}. Recent computational studies have provided detailed insights into the vibrational structure and hydrogen-bond dynamics of water clusters. These include ab initio path integral molecular dynamics simulations that explicitly treat nuclear quantum effects such as proton tunneling and proton delocalization~\cite{Mei1998}, semiclassical dynamics calculations capturing anharmonic effects and mode couplings~\cite{Ceotto2017, Di2018}, and comprehensive reviews of nuclear quantum effects in water~\cite{Ceriotti2016}. Because of the curse of dimensionality, studying the hydrogen-bonded model presents a significant challenge to computational mathematics.

	The cavity quantum electrodynamics (QED) models~\cite{Rabi1936, Rabi1937, Jaynes1963, Tavis1968} constitute a significant contribution to this paper. They provide a novel scientific framework for investigating light--matter interaction. The QED model is a stable and reliable theoretical model, and a substantial body of work has been built around it in recent years, covering quantum gates~\cite{Dull2021}, quantum many-body phenomena~\cite{Smith2021}, the hydrogen molecule and its cation~\cite{Ozhigov2021, Miao2023, MiaoOzhigov2023}, entropy and quantum discord~\cite{MiaoHuihui2024, MiaoLi2025}, and dark states~\cite{Lee1999, Andre2002, Poltl2012, Tanamoto2012, Hansom2014, Ozhigov2020}. In our paper, the hydrogen bond is described by a modified version of the Tavis--Cummings--Hubbard model~\cite{Angelakis2007}, which is readily scaled to complex molecular systems.
	
	The structure of this article is as follows. Sec.~\ref{sec:HBModel} introduces the hydrogen-bonded model, mapping each subsystem to a $\lambda$-type three-level system coupled to two effective phonon modes. Sec.~\ref{sec:Cluster} defines the $[(\mathrm{H}_2\mathrm{O})_2]^m$ cluster and distinguishes it from the $[\mathrm{H}_2\mathrm{O}]^{2m}$ configuration. Sec.~\ref{sec:Coherence} presents the Hamiltonians for the incoherent and coherent cases, where the phonon modes are either independent or shared among subsystems. Sec.~\ref{sec:DMP} describes the density matrix purification method used to efficiently solve the quantum master equation for large $m$. Sec.~\ref{sec:Results} presents the main numerical results, including the dynamics of the cluster, the emergence of dark states, and the effect of phonon inflow on hydrogen bond formation. Sec.~\ref{sec:Conclusion} summarizes the findings and discusses directions for future work.
	
	\section{Hydrogen-bonded model}
	\label{sec:HBModel}
	
	\begin{figure*}
		\centering
        \includegraphics[width=1.\textwidth]{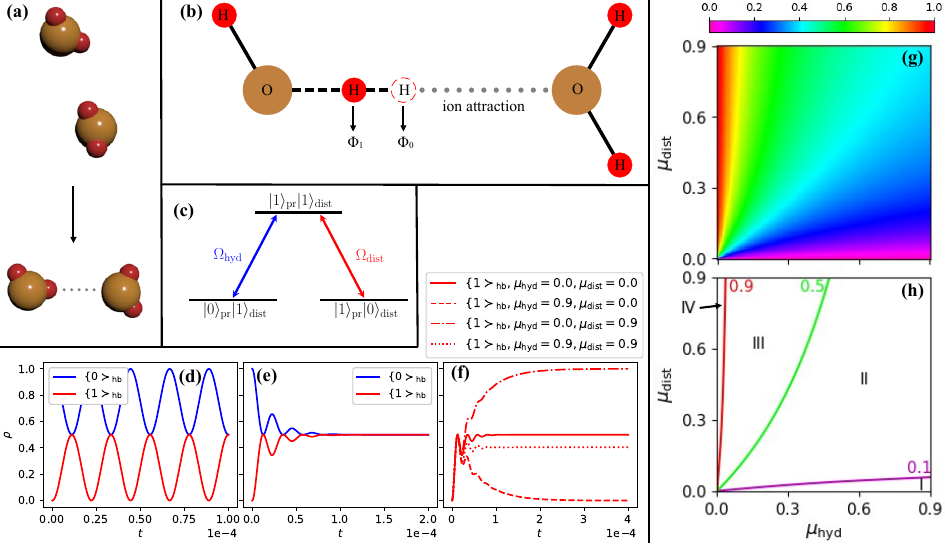}
        \caption{(online color) {\it Hydrogen-bonded model.} The mechanism of hydrogen bond formation is shown in panels (a) and (b). Here, brown circles (balls) represent oxygen atoms, red circles (balls) represent hydrogen atoms, black lines represent covalent bonds, and gray dotted lines represent ion attraction. Panel (c) represents the structure of a $\lambda$-type three-level system. The unitary evolution and dissipative dynamics of the hydrogen-bonded model are obtained in panels (d) and (e), respectively. Panels (f)--(h) display the results under various phonon inflow values. The dividing lines correspond to probabilities of $0.1$ (purple), $0.5$ (green), and $0.9$ (red). Regions I, II, III, and IV correspond to probability intervals $[0,\ 0.1),\ [0.1,\ 0.5),\ [0.5,\ 0.9)$, and $[0.9,\ 1]$, respectively.}
        \label{fig:HBModel}
	\end{figure*}
	
	The target model describing the system of two $\mathrm{H}_2\mathrm{O}$ molecules connected by a hydrogen bond is shown in panels (a) and (b) of Fig.~\ref{fig:HBModel}. The states describing proton micro-vibrations are shown as $\Phi_1$ and $\Phi_0$ in panel (b): $\Phi_1$ corresponds to the proton at the critical point, where it is attracted to the oxygen atom; $\Phi_0$ corresponds to the proton stretched to a distance at which a hydrogen bond is formed and a phonon is released.
	
	The Hamiltonian $H_{\mathrm{hb}}$ describes the system energy as
    \begin{equation}
        H_{\mathrm{hb}}=H_{\mathrm{hyd}}+H_{\mathrm{dist}},
        \label{eq:HamilHBM}
    \end{equation}
    where $H_{\mathrm{hyd}}$ is the Hamiltonian describing the transition of a proton between excited and ground states, corresponding to a micro-vibration. $H_{\mathrm{dist}}$ is the Hamiltonian describing the simplified displacement of a proton in space. This corresponds to a macro-displacement, i.e., the mechanism by which the proton of one water molecule either departs from or approaches the oxygen atom of another water molecule. The rotating wave approximation (RWA) is taken into account~\cite{Wu2007}. $H_{\mathrm{hyd}}$ has the following form:
	\begin{equation}
		\begin{aligned}
        		H_{\mathrm{hyd}}&=\hbar\Omega_{\mathrm{hyd}}a_{\mathrm{hyd}}^{\dag}a_{\mathrm{hyd}}+\hbar\Omega_{\mathrm{hyd}}\sigma_{\mathrm{hyd}}^{\dag}\sigma_{\mathrm{hyd}}\\
        		&+g_{\mathrm{hyd}}\left(a_{\mathrm{hyd}}^{\dag}\sigma_{\mathrm{hyd}}+a_{\mathrm{hyd}}\sigma_{\mathrm{hyd}}^{\dag}\right),
        	\end{aligned}
        \label{eq:HamilHyd}
    \end{equation}
    where $\hbar=h/2\pi$ is the reduced Planck constant, $\Omega_{\mathrm{hyd}}$ is the phonon mode for the proton transition, and $g_{\mathrm{hyd}}$ is the coupling strength between the field (with the annihilation and creation operators $a_{\mathrm{hyd}}$ and $a_{\mathrm{hyd}}^{\dag}$, respectively) and the proton (with the excitation and relaxation operators $\sigma_{\mathrm{hyd}}^{\dag}$ and $\sigma_{\mathrm{hyd}}$, respectively). $H_{\mathrm{dist}}$ has the following form:
    \begin{equation}
    		\begin{aligned}
        		H_{\mathrm{dist}}&=\hbar\Omega_{\mathrm{dist}}a_{\mathrm{dist}}^{\dag}a_{\mathrm{dist}}+\hbar\Omega_{\mathrm{dist}}\sigma_{\mathrm{dist}}^{\dag}\sigma_{\mathrm{dist}}\\
        		&+g_{\mathrm{dist}}\left(a_{\mathrm{dist}}^{\dag}\sigma_{\mathrm{dist}}+a_{\mathrm{dist}}\sigma_{\mathrm{dist}}^{\dag}\right),
        	\end{aligned}
        \label{eq:HamilDist}
    \end{equation}
    where $\Omega_{\mathrm{dist}}$ is the phonon mode associated with proton tunneling, and $g_{\mathrm{dist}}$ is the coupling strength between the field (with the annihilation and creation operators $a_{\mathrm{dist}}$ and $a_{\mathrm{dist}}^{\dag}$, respectively) and the proton (with the tunneling operators $\sigma_{\mathrm{dist}}^{\dag}$ and $\sigma_{\mathrm{dist}}$, respectively). Thus, $H_{\mathrm{hyd}}$ and $H_{\mathrm{dist}}$ are Jaynes--Cummings models, and $H_{\mathrm{hb}}$ is a symmetrical $\lambda$-type three-level system (with $\Omega_{\mathrm{hyd}}=\Omega_{\mathrm{dist}}$), as illustrated in panel (c) of Fig.~\ref{fig:HBModel}.
    
    Although we set $\Omega_{\mathrm{hyd}}=\Omega_{\mathrm{dist}}$ for analytical convenience, the two phonon modes are distinguished by their coupling operators and physical roles. The micro-vibration mode $\Omega_{\mathrm{hyd}}$ couples to the proton's internal state via $\sigma_{\mathrm{hyd}}$ and represents the O--H stretching vibration that modulates the proton's localization within a single potential well. The macro-displacement mode $\Omega_{\mathrm{dist}}$ couples to the proton's center-of-mass coordinate via $\sigma_{\mathrm{dist}}$ and corresponds to the intermolecular donor--acceptor motion (i.e., the O$\cdots$O distance variation), which directly governs hydrogen bond formation. This distinction is supported by extensive studies~\cite{Xantheas1995, Buck2000, Wang2011, Ceotto2017, Di2018} on water clusters: the O--H stretching modes are known to exhibit significant frequency shifts upon hydrogen bond formation, while the donor--acceptor distance is recognized as the key coordinate determining hydrogen bond strength.
    
    The quantum master equation (QME) of an open quantum system under the Markovian approximation has the following form:
    \begin{equation}
		\label{eq:QME}
		\begin{aligned}
			i\hbar\dot{\rho}&=\left[H,\rho\right]+i\hbar\left[\sum_{k\in \mathcal{K}} L_k\left(\rho\right)+\sum_{k'\in \mathcal{K}'} L_{k'}\left(\rho\right)\right]\\
			&=\left[H,\rho\right]+i\hbar\left[\sum_{k\in \mathcal{K}}\gamma_k\left(A_k\rho A_k^{\dag}-\frac{1}{2}\left\{\rho, A_k^{\dag}A_k\right\}\right)\right.\\
			&\left.+\sum_{k'\in \mathcal{K}'}\gamma_{k'}\left(A_k^{\dag}\rho A_k-\frac{1}{2}\left\{\rho, A_kA_k^{\dag}\right\}\right)\right],
		\end{aligned}
	\end{equation}
	where $H$ is the Hamiltonian and $\rho$ is the density matrix. We introduce a graph $\mathcal{K}$ representing the allowed phonon dissipative transitions between states. The edges and vertices of $\mathcal{K}$ represent the permitted dissipations and the states, respectively. $L_k\left(\rho\right)$ is the standard dissipation superoperator corresponding to the jump operator $A_k$ with emission rate $\gamma_k$ for $k\in \mathcal{K}$. Similarly, $\mathcal{K}'$ is a graph representing the allowed photon inflows. $L_{k'}\left(\rho\right)$ is the standard inflow superoperator with inflow rate $\gamma_{k'}$ for $k'\in \mathcal{K}'$. We define the ratio $\mu_k=\frac{\gamma_{k'}}{\gamma_k},\ 0\leq\mu_k<1$.
    
    The Hilbert space takes the following form:
    \begin{equation}
    		|\Psi\rangle_{\mathrm{sys}}=|p_1\rangle_{\Omega_{\mathrm{hyd}}}|p_2\rangle_{\Omega_{\mathrm{dist}}}|l\rangle_{\mathrm{pr}}|d\rangle_{\mathrm{dist}},
    		\label{eq:SpaceHBM}
    \end{equation}
    where $p_1 \in [0,1]$ is the number of phonons in mode $\Omega_{\mathrm{hyd}}$, and $p_2 \in [0,1]$ is the number of phonons in mode $\Omega_{\mathrm{dist}}$. $l=0$ corresponds to the proton in the ground state $\Phi_0$, and $l=1$ corresponds to the proton in the excited state $\Phi_1$. $d=0$ corresponds to the proton being far away, with the two water molecules free. $d=1$ corresponds to the proton being close to the oxygen atom of another water molecule, placing the system at the critical point of hydrogen bond formation. We can divide this Hilbert space into two subspaces. To facilitate the comparison of probabilities associated with subspaces corresponding to different numbers of hydrogen bonds, we denote the subspace with exactly $k$ hydrogen bonds by $\{k\succ_{\mathrm{hb}}$ (instead of the conventional notation $\mathcal{H}_k$). This Hilbert space can thus be divided into two subspaces:
	\begin{subequations}
    		\begin{align}
        		\{0\succ_{\mathrm{hb}} &= \operatorname{span}\{|0\rangle|0\rangle|1\rangle|0\rangle,\ |0\rangle|1\rangle|1\rangle|0\rangle,\ |0\rangle|0\rangle|1\rangle|1\rangle\},\label{eq:State0}\\
        		\{1\succ_{\mathrm{hb}} &= \operatorname{span}\{|1\rangle|0\rangle|0\rangle|1\rangle,\ |0\rangle|0\rangle|0\rangle|1\rangle\},\label{eq:State1}
    		\end{align}
   		\label{eq:States0+1}
	\end{subequations}
	where the basis states are expressed as $|p_1\rangle_{\Omega_{\mathrm{hyd}}}|p_2\rangle_{\Omega_{\mathrm{dist}}}|l\rangle_{\mathrm{pr}}|d\rangle_{\mathrm{dist}}$ according to Eq.~\eqref{eq:SpaceHBM}. According to Eq.~\eqref{eq:States0+1}, the condition for hydrogen bond formation is that the proton is in the ground state, and the probabilities of $\{0\succ_{\mathrm{hb}}$ and $\{1\succ_{\mathrm{hb}}$ depend only on the penultimate qubit $|l\rangle_{\mathrm{pr}}$. For convenience, we assume $|0\rangle_{\mathrm{p+d}}=|0\rangle_{\mathrm{pr}}|1\rangle_{\mathrm{dist}}$, $|1\rangle_{\mathrm{p+d}}=|1\rangle_{\mathrm{pr}}|0\rangle_{\mathrm{dist}}$, and $|2\rangle_{\mathrm{p+d}}=|1\rangle_{\mathrm{pr}}|1\rangle_{\mathrm{dist}}$. Eq.~\eqref{eq:States0+1} can be rewritten as 
	\begin{subequations}
		\begin{align}
			\{0\succ_{\mathrm{hb}} &= \operatorname{span}\{|0\rangle|0\rangle|1\rangle,\ |0\rangle|1\rangle|1\rangle,\ |0\rangle|0\rangle|2\rangle\},\label{eq:NewState0}\\
        		\{1\succ_{\mathrm{hb}} &= \operatorname{span}\{|1\rangle|0\rangle|0\rangle,\ |0\rangle|0\rangle|0\rangle\}.\label{eq:NewState1}
		\end{align}
		\label{eq:NewStates0+1}
	\end{subequations}
	
	We note that this discrete-state criterion for H-bond formation is a coarse-grained representation of the continuous geometric and energetic criteria commonly used in experiments and atomistic simulations. In our model:
	\begin{itemize}
    		\item $|0\rangle_{\mathrm{p+d}}$ corresponds to the proton being localized near the acceptor oxygen atom and in its ground state — this is the H-bonded configuration.
    		\item $|1\rangle_{\mathrm{p+d}}$ corresponds to the proton being far from the acceptor (large O$\cdots$O distance) — a non-bonded configuration.
    		\item $|2\rangle_{\mathrm{p+d}}$ corresponds to the proton being near the acceptor but in an excited (critical) state, representing the transition state just before bond formation — also a non-bonded configuration.
	\end{itemize}
The emission of an $\Omega_{\mathrm{hyd}}$ phonon accompanies the transition from $|2\rangle_{\mathrm{p+d}}$ to $|0\rangle_{\mathrm{p+d}}$, reflecting the coupling between the proton's internal state and the O--H stretching vibration. Similarly, the emission of an $\Omega_{\mathrm{dist}}$ phonon accompanies the transition from $|2\rangle_{\mathrm{p+d}}$ to $|1\rangle_{\mathrm{p+d}}$, reflecting the coupling between proton transfer and the donor--acceptor distance. This coarse-graining enables a tractable quantum-mechanical treatment of coherence effects that are difficult to capture in classical simulations.
    
	The initial state of the hydrogen-bonded model is chosen as $|\Psi(t=0)\rangle=|0\rangle|0\rangle|2\rangle$, and the results are shown in panels (d) and (e) of Fig.~\ref{fig:HBModel}. In both unitary and dissipative evolution, the maximum probability of $\{1\succ_{\mathrm{hb}}$ is 0.5. In the long-time limit, the probabilities of $\{0\succ_{\mathrm{hb}}$ and $\{1\succ_{\mathrm{hb}}$ are each 0.5. The effect of phonon inflow on evolution is studied. In panel (f), phonon inflow with mode $\Omega_{\mathrm{dist}}$ promotes hydrogen bonding, while phonon inflow in mode $\Omega_{\mathrm{hyd}}$ inhibits it. Therefore, it is instructive to study the case where both types of phonon inflow occur simultaneously, as shown in panel (g), where each point on the heat map represents the probability of $\{1\succ_{\mathrm{hb}}$. Panel (h) shows the four dividing lines obtained from panel (g). From these two panels, we see that when $\mu_{\mathrm{hyd}}$ is close to 0, the probability of $\{1\succ_{\mathrm{hb}}$ is greater than or equal to 0.9. In this case, we can assert that a hydrogen bond is formed.
 
	\section{$[(\mathrm{H}_2\mathrm{O})_2]^m\neq[\mathrm{H}_2\mathrm{O}]^{2m}$}
    \label{sec:Cluster}
    
    We use $[(\mathrm{H}_2\mathrm{O})_2]^m$ to define the cluster formed by multiple $(\mathrm{H}_2\mathrm{O})_2$ systems. Here, the water molecules are in pairs; that is, one water molecule is fixed to another via a hydrogen bond. The reason we set it this way is to reduce the complexity of the cluster. If water molecules are not fixed in this way, the $[\mathrm{H}_2\mathrm{O}]^{2m}$ cluster is obtained. We would then need to calculate the spatial motion of all water molecules and the various arrangements and combinations of hydrogen-bonded configurations, which would lead to a dramatic increase in computational cost.
    
    \section{$\mathcal{H}_{\mathrm{incoherent}}^m\neq \mathcal{H}_{\mathrm{coherent}}^m$}
    \label{sec:Coherence}
    
\begin{figure}
		\centering
        \includegraphics[width=.5\textwidth]{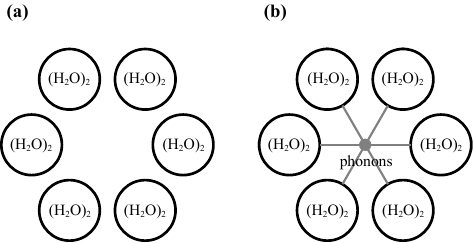} 
        \caption{(online color) {\it Incoherent and coherent cases.} Panel (a) shows the incoherent case, and panel (b) shows the coherent case.}
        \label{fig:Cluster}
    \end{figure}

    In Fig.~\ref{fig:Cluster}, we study two cases of the $[(\mathrm{H}_2\mathrm{O})_2]^m$ cluster: incoherent and coherent. We note that in this work, ``phonon coherence'' refers to quantum coherence mediated by shared vibrational modes, i.e., the ability of the phonon field to create superpositions and interference among different decay channels. This is distinct from the notion of phonon wave coherence in thermal transport, which concerns the phase correlation of lattice waves over extended distances. The Hamiltonian of the incoherent $[(\mathrm{H}_2\mathrm{O})_2]^m$ cluster is given by
    \begin{equation}
    		\begin{aligned}
        		\mathcal{H}_{\mathrm{incoherent}}^m&=\sum_{i=1}^mH_{\mathrm{hb}}^i\\
        		&=\sum_{i=1}^m\left[H_{\mathrm{hyd}}^i+H_{\mathrm{dist}}^i\right]\\
        		&=\sum_{i=1}^m\left[\hbar\Omega_{\mathrm{hyd}}a_{\mathrm{hyd},i}^{\dag}a_{\mathrm{hyd},i}+\hbar\Omega_{\mathrm{hyd}}\sigma_{\mathrm{hyd},i}^{\dag}\sigma_{\mathrm{hyd},i}\right.\\
        		&+g_{\mathrm{hyd}}\left(a_{\mathrm{hyd},i}^{\dag}\sigma_{\mathrm{hyd},i}+a_{\mathrm{hyd},i}\sigma_{\mathrm{hyd},i}^{\dag}\right)\\
        		&+\hbar\Omega_{\mathrm{dist}}a_{\mathrm{dist},i}^{\dag}a_{\mathrm{dist},i}+\hbar\Omega_{\mathrm{dist}}\sigma_{\mathrm{dist},i}^{\dag}\sigma_{\mathrm{dist},i}\\
        		&\left.+g_{\mathrm{dist}}\left(a_{\mathrm{dist},i}^{\dag}\sigma_{\mathrm{dist},i}+a_{\mathrm{dist},i}\sigma_{\mathrm{dist},i}^{\dag}\right)\right],
        \end{aligned}
        \label{eq:HamilIncoherent}
    \end{equation}
    where $m\geq 1$. The Hilbert space takes the following form:
    \begin{equation}
    		|\Psi\rangle_{\mathrm{incoherent}}^m=\bigotimes_{i=1}^m|p_1^i\rangle_{\Omega_{\mathrm{hyd}}}|p_2^i\rangle_{\Omega_{\mathrm{dist}}}|a^i\rangle_{\mathrm{p+d}},
    		\label{eq:SpaceIncoherent}
    \end{equation}
    where $p_1^i,\ p_2^i \in \{0,\ 1\}$, and according to Eq.~\eqref{eq:NewStates0+1}, $a^i\in\{0,\ 1,\ 2\}$.
    
    In the coherent case, the phonons are shared by the entire system. The Hamiltonian of the coherent $[(\mathrm{H}_2\mathrm{O})_2]^m$ cluster is given by
    \begin{equation}
    		\begin{aligned}
        		\mathcal{H}_{\mathrm{coherent}}^m&=\hbar\Omega_{\mathrm{hyd}}a_{\mathrm{hyd}}^{\dag}a_{\mathrm{hyd}}+\hbar\Omega_{\mathrm{dist}}a_{\mathrm{dist}}^{\dag}a_{\mathrm{dist}}\\
        		&+\sum_{i=1}^m\left[\hbar\Omega_{\mathrm{hyd}}\sigma_{\mathrm{hyd},i}^{\dag}\sigma_{\mathrm{hyd},i}+\hbar\Omega_{\mathrm{dist}}\sigma_{\mathrm{dist},i}^{\dag}\sigma_{\mathrm{dist},i}\right.\\
        		&+g_{\mathrm{hyd}}\left(a_{\mathrm{hyd}}^{\dag}\sigma_{\mathrm{hyd},i}+a_{\mathrm{hyd}}\sigma_{\mathrm{hyd},i}^{\dag}\right)\\
        		&\left.+g_{\mathrm{dist}}\left(a_{\mathrm{dist}}^{\dag}\sigma_{\mathrm{dist},i}+a_{\mathrm{dist}}\sigma_{\mathrm{dist},i}^{\dag}\right)\right],
        \end{aligned}
        \label{eq:HamilСoherent}
    \end{equation}
    where $m\geq 2$. The Hilbert space takes the following form:
	\begin{equation}
    		|\Psi\rangle_{\mathrm{coherent}}^m=|p_1\rangle_{\Omega_{\mathrm{hyd}}}|p_2\rangle_{\Omega_{\mathrm{dist}}}\bigotimes_{i=1}^m|a^i\rangle_{\mathrm{p+d}},
    		\label{eq:SpaceCoherent}
    \end{equation}
    where $p_1,\ p_2 \in \{0,\ 1,\ \dots,\ m\}$.
     
	For the incoherent case, the subspace with exactly $k$ hydrogen bonds is the direct sum of tensor products of the single-subsystem subspaces:
	\begin{equation}
		\{k\succ_{\mathrm{hb}}^m = \bigoplus_{k_1+\cdots+k_m=k} \bigotimes_{i=1}^m \{k_i\succ_{\mathrm{hb}},
		\label{eq:StatesIncoherent}
	\end{equation}
	where $m\geq 1$, $k\in[0,m]$, $k_i\in\{0,1\}$. Here $\{k_i\succ_{\mathrm{hb}}$ denotes the subspace of the $i$-th subsystem with $k_i$ hydrogen bonds, and $\bigoplus$ indicates the direct sum over all combinations satisfying the constraint. For the coherent case, the subspace with exactly $k$ hydrogen bonds has the following form:
	\begin{equation}
		\{k\succ_{\mathrm{hb}}^m = \operatorname{span}\left\{ |p_1\rangle_{\Omega_{\mathrm{hyd}}}|p_2\rangle_{\Omega_{\mathrm{dist}}} \bigotimes_{i=1}^m |a^i\rangle_{\mathrm{p+d}} \; \middle| \; \sum_{i=1}^m \delta_{a^i0} = k \right\},
		\label{eq:StatesCoherent}
	\end{equation}
	where $m\geq 2$, $k\in[0,m]$, $a^i\in\{0,1,2\}$, and $\delta_{a^i0}$ is the Kronecker delta:
    \begin{equation}
    		\delta_{a^i0}=
    		\begin{cases}
    			1,\ a^i=0,\\
    			0,\ a^i\neq0.
    		\end{cases}
    		\label{eq:Kronecker}
    \end{equation}
  
    \section{Density matrix purification}
    \label{sec:DMP}
    
    To solve the QME, we use an iteration scheme consisting of two steps:
    \begin{subequations}
		\begin{align}
			\widetilde{\rho}(t+\Delta t)&=\exp\left(-\frac{i}{\hbar}H\Delta t\right)\rho(t)\exp\left(\frac{i}{\hbar}H\Delta t\right),\label{equation:Eulermethod1}\\
			\rho(t+\Delta t)&=\widetilde{\rho}(t+\Delta t)+\mathcal{L}(\widetilde{\rho}(t+\Delta t))\Delta t,\label{equation:Eulermethod2}
		\end{align}
		\label{equation:Eulermethod}
	\end{subequations}
	where $\Delta t$ is the time step.
	
	For a $[(\mathrm{H}_2\mathrm{O})_2]^m$ cluster, as $m$ increases, the dimension of the density matrix also increases, in both the incoherent and coherent cases. In particular, when $m\geq 6$, the dimension of the density matrix exceeds the capacity of a typical CPU. Some supercomputing algorithms~\cite{LiMiao2024, MiaoOzhigov2024} have been proposed to solve this problem caused by the large dimensionality. In this paper, we do not use parallel computing algorithms to split the density matrix but instead adopt a new method --- density matrix purification (DMP), which takes advantage of a property of open quantum systems: there is no unitary evolution between systems with different energies. For example, the energy of the density matrix $\rho_A$ is always $E_A$. When a phonon with energy $\hbar\Omega$ escapes from the system described by $\rho_A$, another density matrix $\rho_B$ with energy $E_B$ is formed, and $E_B=E_A-\hbar\Omega$. The process from $\rho_A$ to $\rho_B$ is a dissipative process. Based on this, we can divide the density matrix of the $[(\mathrm{H}_2\mathrm{O})_2]^m$ cluster system into many small blocks according to different energies and types. Each block represents a density matrix. Unitary evolution occurs only inside each block. The blocks interact with each other only through the QME. In practice, we apply the DMP to the density matrix when $m=6$ and compare it with the original case (see Tab.~\ref{tab:DMP}): the purified blocks require much less memory than the original density matrix, especially in the incoherent case.
	 \begin{table}[!htpb]
        \centering
		\begin{tabular}{|c|c|c|c|c|}
			\hline
			& org\_dim & num\_bls & max\_dim\_bl & $\frac{\sum_i\mathrm{dim\_bl\_i}\times\mathrm{\dim\_bl\_i}}{\mathrm{org\_dim}\times\mathrm{org\_dim}}$ \\
			\hline
			incoherent case & 15625 & 729 & 729 & 0.726\% \\
            	\hline
			coherent case & 6075 & 28 & 729 & 7.739\% \\
            	\hline
		\end{tabular}
		\caption{{\it Comparison of the DMP case with the original case.} Here, org\_dim is the dimension of the original density matrix, num\_bls is the number of blocks in the DMP case, max\_dim\_bl is the maximum block dimension, and $\frac{\sum_i\mathrm{dim\_bl\_i}\times\mathrm{\dim\_bl\_i}}{\mathrm{org\_dim}\times\mathrm{org\_dim}}$ is the ratio of memory occupied in the DMP case to that in the original case.}
		\label{tab:DMP}
    \end{table}
    
    \section{Main results}
    \label{sec:Results}
    
    \begin{figure*}
        \centering
        \includegraphics[width=1.\textwidth]{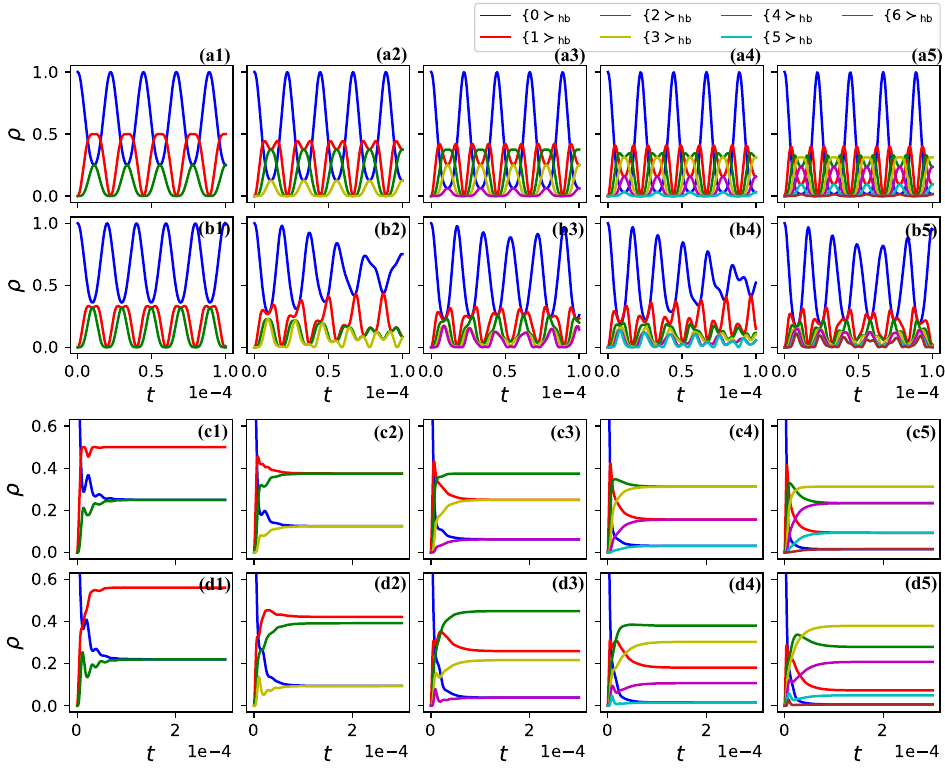}
        \caption{(online color) {\it Evolution of $[(\mathrm{H}_2\mathrm{O})_2]^m$ cluster.} The first to fifth columns correspond to $m=2$ to $6$, respectively. Rows (a) and (b): unitary evolution in the incoherent and coherent cases. The oscillations reflect Rabi exchange between the proton and the shared phonon modes; the period is set by $g_{\mathrm{hyd}}=g_{\mathrm{dist}}$. In the coherent case (b), the oscillations deform with increasing $m$ due to collective coupling among subsystems. Rows (c) and (d): dissipative evolution in the incoherent and coherent cases. In the coherent case (d), a slow decay component appears for larger $m$, corresponding to the lifetime of dark-state populations. The steady-state redistribution toward intermediate H-bond counts (row (d)) is a signature of coherence-induced dark-state protection.}
        \label{fig:UnitaryDissipative}
    \end{figure*}
    
    In contrast to the single hydrogen-bonded system in Fig.~\ref{fig:HBModel}, here we study the $[(\mathrm{H}_2\mathrm{O})_2]^m$ cluster composed of multiple hydrogen-bonded systems, with $m\in[2,\ 6]$. First, we study the unitary evolutions in a closed quantum system. Rows (a) and (b) of Fig.~\ref{fig:UnitaryDissipative} show oscillations in all panels. In the incoherent case (row (a)), the oscillations are perfectly periodic because each hydrogen-bonded system is independent. However, when phonon coherence is established (row (b)), the oscillations are deformed. In particular, the larger the $m$, the more obvious the deformation, because the coherence becomes stronger. In an open system, rows (c) and (d) of Fig.~\ref{fig:UnitaryDissipative} show that oscillations gradually decay, and each state eventually settles to a stable value. In row (c), the results show that the final stable value of each state (see Tab.~\ref{tab:Incoherent}) satisfies the following equation: $\rho\left(\{k\succ_{\mathrm{hb}}^m\right)=\frac{C_m^k}{2^m}$. In row (d), phonon coherence prevents the stable value of each state from satisfying $\frac{C_m^k}{2^m}$. We compare the stable values of each state in row (c) with those in row (d) to obtain Tab.~\ref{tab:Comparison}. From Tab.~\ref{tab:Comparison}, we see that for $m\in[2,\ 6]$, the probabilities of the states in the middle of each column increase due to the coherence effect, while those on both sides decrease. This redistribution will be explained in detail later, in conjunction with the analysis of dark states.

    \begin{table}[!htpb]
        \centering
		\begin{tabular}{|c|c|c|c|c|c|c|}
			\hline
			& $m=1$ & $m=2$ & $m=3$ & $m=4$ & $m=5$ & $m=6$ \\
			\hline
			$\{0\succ_{\mathrm{hb}}^m$ & $\frac{C_1^0}{2^1}$ & $\frac{C_2^0}{2^2}$ & $\frac{C_3^0}{2^3}$ & $\frac{C_4^0}{2^4}$ & $\frac{C_5^0}{2^5}$ & $\frac{C_6^0}{2^6}$ \\
            	\hline
			$\{1\succ_{\mathrm{hb}}^m$ & $\frac{C_1^1}{2^1}$ & $\frac{C_2^1}{2^2}$ & $\frac{C_3^1}{2^3}$ & $\frac{C_4^1}{2^4}$ & $\frac{C_5^1}{2^5}$ & $\frac{C_6^1}{2^6}$ \\
            	\hline
            	$\{2\succ_{\mathrm{hb}}^m$ & & $\frac{C_2^2}{2^2}$ & $\frac{C_3^2}{2^3}$ & $\frac{C_4^2}{2^4}$ & $\frac{C_5^2}{2^5}$ & $\frac{C_6^2}{2^6}$ \\
            	\hline
            	$\{3\succ_{\mathrm{hb}}^m$ & & & $\frac{C_3^3}{2^3}$ & $\frac{C_4^3}{2^4}$ & $\frac{C_5^3}{2^5}$ & $\frac{C_6^3}{2^6}$ \\
            	\hline
            	$\{4\succ_{\mathrm{hb}}^m$ & & & & $\frac{C_4^4}{2^4}$ & $\frac{C_5^4}{2^5}$ & $\frac{C_6^4}{2^6}$ \\
            	\hline
            	$\{5\succ_{\mathrm{hb}}^m$ & & & & & $\frac{C_5^5}{2^5}$ & $\frac{C_6^5}{2^6}$ \\
            	\hline
            	$\{6\succ_{\mathrm{hb}}^m$ & & & & & & $\frac{C_6^6}{2^6}$ \\
			\hline
		\end{tabular}
		\caption{{\it Expected probabilities in the incoherent case.}}
		\label{tab:Incoherent}
    \end{table}
    
    \begin{table}[!htpb]
        \centering
		\begin{tabular}{|c|c|c|c|c|c|c|}
			\hline
			& $m=1$ & $m=2$ & $m=3$ & $m=4$ & $m=5$ & $m=6$ \\
			\hline
			$\{0\succ_{\mathrm{hb}}^m$ & & $-$ & $-$ & $-$ & $-$ & $-$ \\
            	\hline
			$\{1\succ_{\mathrm{hb}}^m$ & & + & + & + & + & $-$ \\
            	\hline
            	$\{2\succ_{\mathrm{hb}}^m$ & & $-$ & + & + & + & + \\
            	\hline
            	$\{3\succ_{\mathrm{hb}}^m$ & & & $-$ & $-$ & $-$ & + \\
            	\hline
            	$\{4\succ_{\mathrm{hb}}^m$ & & & & $-$ & $-$ & $-$ \\
            	\hline
            	$\{5\succ_{\mathrm{hb}}^m$ & & & & & $-$ & $-$ \\
            	\hline
            	$\{6\succ_{\mathrm{hb}}^m$ & & & & & & $-$ \\
			\hline
		\end{tabular}
		\caption{{\it Comparison of the coherent case with the incoherent case.} The symbols ``$+$'' and ``$-$'' indicate that the probability in the coherent case is greater or smaller, respectively, than that in the incoherent case.}
		\label{tab:Comparison}
    \end{table}
    
    	\begin{figure}
        \centering
        \includegraphics[width=.5\textwidth]{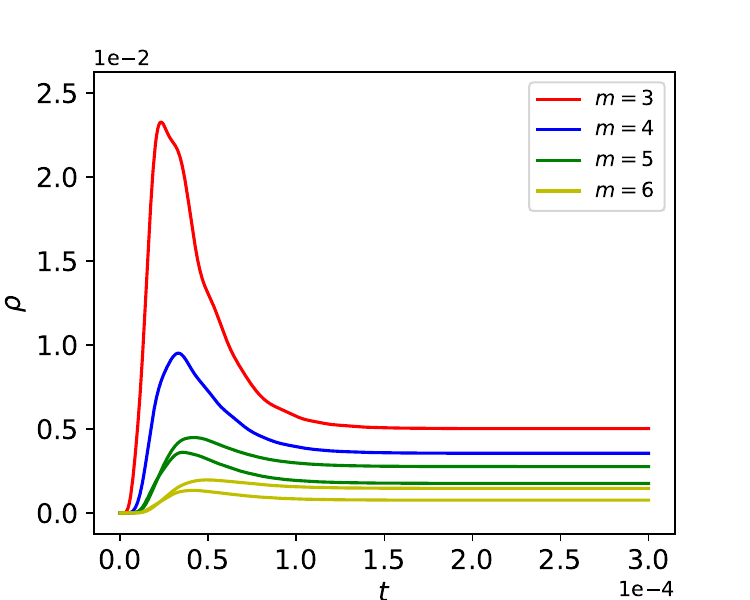}
        \caption{(online color) {\it The dark states for $m\geq 3$.} These states arise from destructive interference in the coupling of multiple subsystems to the shared phonon modes and are defined by $\bar{\sigma}_m |D\rangle = 0$. For $m=3$, the dark state $|D\rangle_{123}$ (Eq.~\eqref{eq:D_BasisM=3}) is a superposition of six basis states, each containing exactly one subsystem in the excited state $|2\rangle$ and the remaining two in the ground states $|0\rangle$ or $|1\rangle$. For $m=4$, the dark-state manifold includes product states derived from $|D\rangle_{123}$ (Eq.~\eqref{eq:DarkStatesM=4}) as well as intrinsically four-body dark states (Eq.~\eqref{eq:DarkStateM=4Plus}). For $m>4$, additional dark state structures emerge. The distribution of dark states across different H-bond sectors underlies the steady-state redistribution shown in Fig.~\ref{fig:UnitaryDissipative} (d).}
        \label{fig:DarkStates}
    \end{figure}
    
	Consideration is also given to the dark states arising from the phonon coherence. The results of Fig.~\ref{fig:DarkStates} show that when $m\geq 3$, dark states appear in the coherent case. When $m=3$, the dark state is a superposition of six basis states, each of which reaches a non-zero stable value. For the coherent $[(\mathrm{H}_2\mathrm{O})_2]^3$ cluster, the dark state takes the following form:
    \begin{equation}
    		|\mathrm{D}\rangle_{\mathrm{coherent}}^3=|00\rangle\otimes|\mathrm{D_{basis}}\rangle_{123},
    		\label{eq:DarkStateM=3}    		
	\end{equation}
	with
	\begin{equation}
    		|\mathrm{D_{basis}}\rangle_{123}=\left(|120\rangle-|102\rangle-|210\rangle+|201\rangle+|012\rangle-|021\rangle\right)_{123},
    		\label{eq:D_BasisM=3}    		
	\end{equation}
	where the subscripts $1,2,3$ label the three hydrogen-bonded subsystems. The operator $\bar{\sigma}^m$ represents the sum of all relaxation operators:
	\begin{equation}
		\bar{\sigma}^m=\sum_{i=1}^m\left(\sigma_{\mathrm{hyd},i}+\sigma_{\mathrm{dist},i}\right).
		\label{eq:AllSigma}
	\end{equation}
	One can verify that $\bar{\sigma}^3|\mathrm{D}\rangle_{\mathrm{coherent}}^3=0$, confirming that this is indeed a dark state. The validation procedures for all dark states presented in this paper are detailed in the \href{https://drive.google.com/drive/folders/125Pl9yFLYVTrlZHtbqqSgSp4tWsVwbyf?usp=sharing}{Supplementary Information}. Thus, based on Eqs.~\eqref{eq:DarkStateM=3} and \eqref{eq:D_BasisM=3}, we can obtain the dark states when $m\geq4$. For example, when $m=4$, the dark states are given by
	\begin{subequations}
    		\begin{align}
    			|\mathrm{D}1\rangle_{\mathrm{coherent}}^4&=|00\rangle\otimes|\mathrm{D_{basis}}\rangle_{123}\otimes|0\rangle_{4},\label{eq:DarkStatesM=4_1}\\
    			|\mathrm{D}2\rangle_{\mathrm{coherent}}^4&=|00\rangle\otimes|\mathrm{D_{basis}}\rangle_{124}\otimes|0\rangle_{3},\label{eq:DarkStatesM=4_2}\\
    			|\mathrm{D}3\rangle_{\mathrm{coherent}}^4&=|00\rangle\otimes|\mathrm{D_{basis}}\rangle_{134}\otimes|0\rangle_{2},\label{eq:DarkStatesM=4_3}\\
    			|\mathrm{D}4\rangle_{\mathrm{coherent}}^4&=|00\rangle\otimes|\mathrm{D_{basis}}\rangle_{234}\otimes|0\rangle_{1},\label{eq:DarkStatesM=4_4}\\
    			|\mathrm{D}5\rangle_{\mathrm{coherent}}^4&=|00\rangle\otimes|\mathrm{D_{basis}}\rangle_{123}\otimes|1\rangle_{4},\label{eq:DarkStatesM=4_5}\\
    			|\mathrm{D}6\rangle_{\mathrm{coherent}}^4&=|00\rangle\otimes|\mathrm{D_{basis}}\rangle_{124}\otimes|1\rangle_{3},\label{eq:DarkStatesM=4_6}\\
    			|\mathrm{D}7\rangle_{\mathrm{coherent}}^4&=|00\rangle\otimes|\mathrm{D_{basis}}\rangle_{134}\otimes|1\rangle_{2},\label{eq:DarkStatesM=4_7}\\
    			|\mathrm{D}8\rangle_{\mathrm{coherent}}^4&=|00\rangle\otimes|\mathrm{D_{basis}}\rangle_{234}\otimes|1\rangle_{1},\label{eq:DarkStatesM=4_8}
    		\end{align}
    		\label{eq:DarkStatesM=4}    		
	\end{subequations}
	where $|0\rangle$ and $|1\rangle$ denote the ground states. Thus, for $m\geq3$, the dark states are constructed from the three basic elements $|\mathrm{D_{basis}}\rangle$, $|0\rangle$, and $|1\rangle$. However, for $m=4$, we find three additional dark states that cannot be constructed from these three basic elements, or more precisely, cannot be obtained from Eq.~\eqref{eq:D_BasisM=3}. These dark states have the following form:
	\begin{subequations}
		\begin{align}
    			|\mathrm{D}9\rangle_{\mathrm{coherent}}^4=|00\rangle\otimes|\mathrm{D_{basis}}\rangle_{1234},\label{eq:DarkStatesM=4_9}\\
    			|\mathrm{D}10\rangle_{\mathrm{coherent}}^4=|00\rangle\otimes|\mathrm{D_{basis}'}\rangle_{1234},\label{eq:DarkStatesM=4_10}\\
    			|\mathrm{D}11\rangle_{\mathrm{coherent}}^4=|00\rangle\otimes|\mathrm{D_{basis}''}\rangle_{1234},\label{eq:DarkStatesM=4_11}
		\end{align}
		\label{eq:DarkStateM=4Plus}
	\end{subequations}
	where
	\begin{subequations}
    		\begin{align}
    			|\mathrm{D_{basis}}\rangle_{1234}&=\left(|1210\rangle-|1201\rangle-|1012\rangle+|1021\rangle\right.\nonumber\\
    			&\left.-|2110\rangle+|2101\rangle+|0112\rangle-|0121\rangle\right)_{1234},\label{eq:D_BasisM=4}\\
    			|\mathrm{D_{basis}'}\rangle_{1234}&=\left(|1020\rangle-|1002\rangle-|2010\rangle+|2001\rangle\right.\nonumber\\
    			&\left.-|0120\rangle+|0102\rangle+|0210\rangle-|0201\rangle\right)_{1234},\label{eq:D_BasisM=4'}\\
    			|\mathrm{D_{basis}''}\rangle_{1234}&=\left(|1120\rangle-|1102\rangle+|1200\rangle-|2100\rangle\right.\nonumber\\
    			&-|2011\rangle+|0211\rangle-|0012\rangle+|0021\rangle\nonumber\\
    			&-|1210\rangle+|1012\rangle-|1020\rangle+|2101\rangle\nonumber\\
    			&\left.+|2010\rangle-|0121\rangle+|0102\rangle-|0201\rangle\right)_{1234}.\label{eq:D_BasisM=4''}   
    		\end{align}	
	\end{subequations}
	
	\begin{figure*}
        \centering
        \includegraphics[width=.7\textwidth]{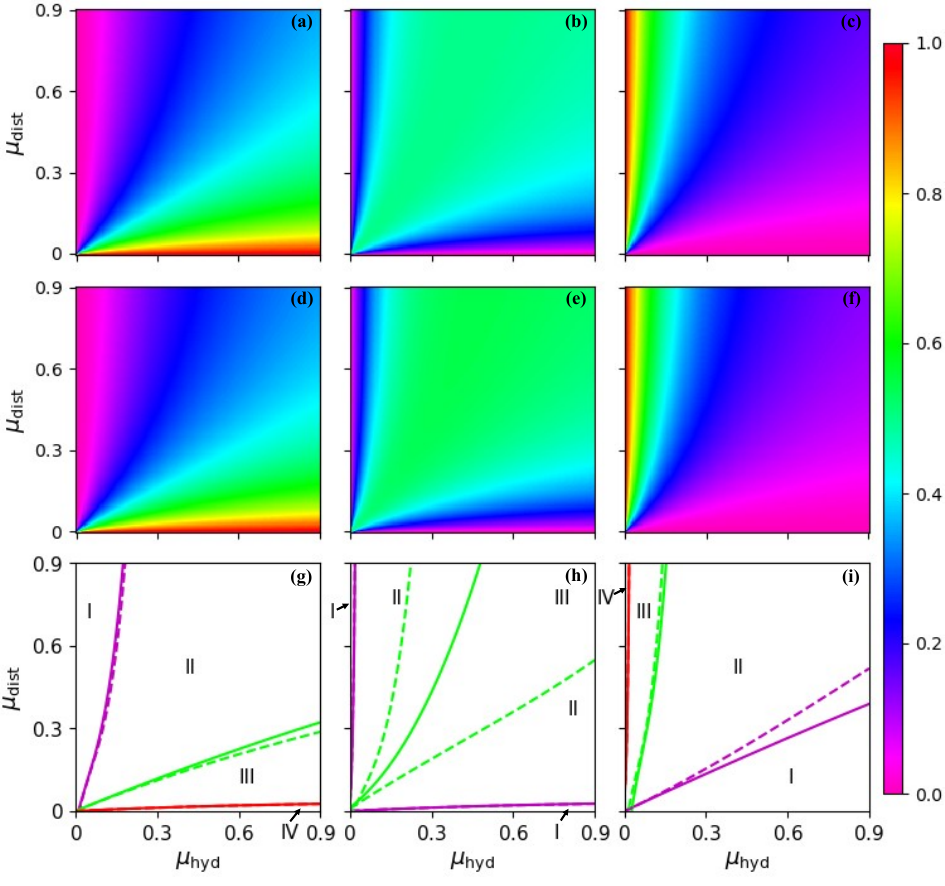}
        \caption{(online color) {\it Effect of inflow intensities $\mu_{\mathrm{hyd}}$ and $\mu_{\mathrm{dist}}$ on the dynamics of the $[(\mathrm{H}_2\mathrm{O})_2]^2$ cluster.} The first, second, and third columns correspond to $\{0\succ_{\mathrm{hb}}^2$, $\{1\succ_{\mathrm{hb}}^2$, and $\{2\succ_{\mathrm{hb}}^2$, respectively. The first row corresponds to the incoherent case, and the second row corresponds to the coherent case. The third row merges the dividing lines from the above two rows into the same coordinate system. Solid lines correspond to the incoherent case, and dashed lines correspond to the coherent case.}
        \label{fig:EffectInflows}
    \end{figure*}
	
	The dark states identified above have a clear physical interpretation. They arise from destructive interference in the collective coupling to the shared phonon modes, effectively decoupling these states from dissipation. Their existence is rooted in the permutation symmetry of the system. The Hamiltonian and dissipative channels are invariant under any exchange of the $m$ subsystems, and dark states are defined by $\bar{\sigma}^m|\mathrm{D}\rangle=0$. For $m=1$ and $m=2$, the Hilbert space does not contain nontrivial solutions to this condition; in particular, for $m=2$, only subradiant (long-lived) states exist. For $m=3$, a fully antisymmetric combination (Eq.~\eqref{eq:DarkStateM=3}) satisfies $\bar{\sigma}^3|\mathrm{D}\rangle=0$, giving a true dark state. For $m>3$, dark states can be constructed by tensoring the $m=3$ dark state with additional subsystems in the ground state, since ground states do not contribute to $\bar{\sigma}^m$. Moreover, for $m=4$, new dark states (Eqs.~\eqref{eq:DarkStateM=4Plus}) appear that are not of this product form, indicating that the dark-state manifold becomes increasingly rich as $m$ grows.
	
	Revisiting Tab.~\ref{tab:Comparison}, we see that this redistribution can be understood through collective interference effects caused by shared phonon modes. For $m=2$, the antisymmetric combination $|\psi_-\rangle_{12}=(|01\rangle_{12}-|10\rangle_{12})/\sqrt{2}$ in the $\{1\succ_{\mathrm{hb}}$ sector is subradiant --- its decay via phonon emission is strongly suppressed due to destructive interference --- while the symmetric combinations $|00\rangle_{12}$ and $|11\rangle_{12}$ in the $\{0\succ_{\mathrm{hb}}$ and $\{2\succ_{\mathrm{hb}}$ sectors undergo fast coherent exchange. This leads to population accumulation in $\{1\succ_{\mathrm{hb}}$ at the expense of the edge sectors. For $m=3$, both $\{1\succ_{\mathrm{hb}}$ and $\{2\succ_{\mathrm{hb}}$ contain subradiant states. In addition, a true dark state (Eq.~\eqref{eq:DarkStateM=3}) emerges exclusively in the $\{1\succ_{\mathrm{hb}}$ sector, providing extra population trapping. Consequently, both intermediate sectors gain population, while $\{0\succ_{\mathrm{hb}}$ and $\{3\succ_{\mathrm{hb}}$ lose population. For $m\ge 4$, subradiant states are most abundant in the middle sectors ($k \approx m/2$), and dark states are distributed asymmetrically across sectors, with a tendency to favor the side closer to $\{0\succ_{\mathrm{hb}}$. This asymmetry, combined with total probability conservation, results in a ``squeezed'' steady-state distribution that is shifted slightly toward the $\{0\succ_{\mathrm{hb}}$ side, as reflected in Tab.~\ref{tab:Comparison}.
	
	Finally, the effect of phonon inflow on hydrogen bond formation is investigated. Results for the $[(\mathrm{H}_2\mathrm{O})_2]^2$ cluster are shown in Fig.~\ref{fig:EffectInflows}: the larger $\mu_{\mathrm{hyd}}$ is and the smaller $\mu_{\mathrm{dist}}$ is, the larger the value of $\{0\succ_{\mathrm{hb}}^2$; conversely, the smaller $\mu_{\mathrm{hyd}}$ is and the larger $\mu_{\mathrm{dist}}$ is, the larger the value of $\{2\succ_{\mathrm{hb}}^2$. Thus, phonon inflow with mode $\Omega_{\mathrm{dist}}$ promotes hydrogen bond formation, while phonon inflow with mode $\Omega_{\mathrm{hyd}}$ prevents hydrogen bond formation. The third row shows that the effect of phonon coherence is relatively insignificant compared with the effect of phonon inflow on hydrogen bond formation. In addition, when $m>2$, we obtain similar results, which can be found in the \href{https://drive.google.com/drive/folders/125Pl9yFLYVTrlZHtbqqSgSp4tWsVwbyf?usp=sharing}{Supplementary Information}.

    \section{Concluding discussion and future work}
    \label{sec:Conclusion}
    
    We propose a simplified open-quantum-system model for hydrogen-bond formation in water clusters, where each subsystem is mapped to a $\lambda$-type three-level system coupled to two effective phonon modes: a micro-vibration mode ($\Omega_{\mathrm{hyd}}$) representing the O--H stretching vibration, and a macro-displacement mode ($\Omega_{\mathrm{dist}}$) representing the intermolecular donor--acceptor motion. We study the $[(\mathrm{H}_2\mathrm{O})_2]^m$ cluster for $m=2$ to $6$ in the incoherent case (independent phonon modes) and the coherent case (shared phonon modes). Our results reveal that phonon coherence significantly alters the dynamics. In the dissipative case, coherence induces a redistribution of steady-state populations: intermediate H-bond counts are enhanced while edge counts are suppressed --- a ``squeezing'' effect explained by the interplay of subradiant states and dark states. For $m\ge 3$, true dark states emerge, rooted in the permutation symmetry of the system. We also investigate the effect of phonon inflow: inflow of $\Omega_{\mathrm{dist}}$ phonons promotes hydrogen bond formation, while inflow of $\Omega_{\mathrm{hyd}}$ phonons inhibits it. The effect of phonon coherence on the steady-state distribution is relatively weak compared to that of phonon inflow.

	Our work complements existing computational studies of water clusters, such as ab initio path integral molecular dynamics, semiclassical dynamics, and studies of nuclear quantum effects. While these approaches provide atomistic details and accurate potential energy surfaces, they typically treat nuclear motion classically or semiclassically and do not systematically address quantum coherence effects. Our open-quantum-system perspective offers a minimal yet analytically tractable framework to study these coherent effects. Although the model is intentionally simplified, its predictions could be tested against ab initio molecular dynamics or path integral simulations by coarse-graining atomistic data to map onto our discrete-state description, representing a promising direction for future research. The methods introduced here provide a foundation for extending the framework to more complex systems.

    \begin{acknowledgments}
	The study was funded by the China Scholarship Council, project number 202108090483. The author acknowledges the Center for Collective Use of Ultra-High-Performance Computing Resources (\url{https://parallel.ru/}) at Lomonosov Moscow State University for providing supercomputer resources that contributed to the research results.
    \end{acknowledgments}

	\bibliography{bibliography}

@article{McArdle2020,
  title = {Quantum computational chemistry},
  author = {McArdle, Sam and Endo, Suguru and Aspuru-Guzik, Al\'an and Benjamin, Simon C. and Yuan, Xiao},
  journal = {Rev. Mod. Phys.},
  volume = {92},
  issue = {1},
  pages = {015003},
  numpages = {51},
  year = {2020},
  month = {Mar},
  publisher = {American Physical Society},
  doi = {10.1103/RevModPhys.92.015003},
  url = {https://link.aps.org/doi/10.1103/RevModPhys.92.015003}
}

@article{Baiardi2023,
author = {Baiardi, Alberto and Christandl, Matthias and Reiher, Markus},
title = {Quantum Computing for Molecular Biology},
journal = {ChemBioChem},
volume = {24},
number = {13},
pages = {e202300120},
doi = {https://doi.org/10.1002/cbic.202300120},
url = {https://chemistry-europe.onlinelibrary.wiley.com/doi/abs/10.1002/cbic.202300120},
year = {2023}
}

@book{Albuquerque2021,
title = {Quantum Chemistry Simulation of Biological Molecules},
publisher = {Cambridge University Press},
year = {2021},
author = {Albuquerque, EL and Fulco, UL and Caetano, EWS and Freire, VN},
}

@Article{Moore1912,
author ="Moore, Tom Sidney and Winmill, Thomas Field",
title  ="{CLXXVII}.---The state of amines in aqueous solution",
journal  ="J. Chem. Soc.{,} Trans.",
year  ="1912",
volume  ="101",
issue  ="0",
pages  ="1635-1676",
publisher  ="The Royal Society of Chemistry",
doi  ="10.1039/CT9120101635",
url  ="http://dx.doi.org/10.1039/CT9120101635"}

@article{Latimer1920,
author = {Latimer, Wendell M. and Rodebush, Worth H.},
title = {POLARITY AND IONIZATION FROM THE STANDPOINT OF THE LEWIS THEORY OF VALENCE.},
journal = {Journal of the American Chemical Society},
volume = {42},
number = {7},
pages = {1419-1433},
year = {1920},
doi = {10.1021/ja01452a015},
URL = {https://doi.org/10.1021/ja01452a015}
}

@article{Ignacio2023,
author = {Ignacio Gustin  and Chang Woo Kim  and David W. McCamant  and Ignacio Franco },
title = {Mapping electronic decoherence pathways in molecules},
journal = {Proceedings of the National Academy of Sciences},
volume = {120},
number = {49},
pages = {e2309987120},
year = {2023},
doi = {10.1073/pnas.2309987120},
URL = {https://www.pnas.org/doi/abs/10.1073/pnas.2309987120}}

@article{He2022,
  title = {Observation of Biradical Spin Coupling through Hydrogen Bonds},
  author = {He, Yang and Li, Na and Castelli, Ivano E. and Li, Ruoning and Zhang, Yajie and Zhang, Xue and Li, Chao and Wang, Bingwu and Gao, Song and Peng, Lianmao and Hou, Shimin and Shen, Ziyong and L\"u, Jing-Tao and Wu, Kai and Hedeg\aa{}rd, Per and Wang, Yongfeng},
  journal = {Phys. Rev. Lett.},
  volume = {128},
  issue = {23},
  pages = {236401},
  numpages = {6},
  year = {2022},
  month = {Jun},
  publisher = {American Physical Society},
  doi = {10.1103/PhysRevLett.128.236401},
  url = {https://link.aps.org/doi/10.1103/PhysRevLett.128.236401}
}

@article{Danko2022,
title = {Thermal stability of solitons in protein $\alpha$-helices},
journal = {Chaos, Solitons \& Fractals},
volume = {155},
pages = {111644},
year = {2022},
issn = {0960-0779},
doi = {https://doi.org/10.1016/j.chaos.2021.111644},
url = {https://www.sciencedirect.com/science/article/pii/S096007792100998X},
author = {Danko D. Georgiev and James F. Glazebrook}
}

@article{Di2018,
    author = {Di Liberto, Giovanni and Conte, Riccardo and Ceotto, Michele},
    title = "{“Divide-and-conquer” semiclassical molecular dynamics: An application to water clusters}",
    journal = {The Journal of Chemical Physics},
    volume = {148},
    number = {10},
    pages = {104302},
    year = {2018},
    month = {03},
    issn = {0021-9606},
    doi = {10.1063/1.5023155},
    url = {https://doi.org/10.1063/1.5023155},
}

@article{Yamada2020,
  title = {Quantum valence bond ice theory for proton-driven quantum spin-dipole liquids},
  author = {Yamada, Masahiko G. and Tada, Yasuhiro},
  journal = {Phys. Rev. Res.},
  volume = {2},
  issue = {4},
  pages = {043077},
  numpages = {8},
  year = {2020},
  month = {Oct},
  publisher = {American Physical Society},
  doi = {10.1103/PhysRevResearch.2.043077},
  url = {https://link.aps.org/doi/10.1103/PhysRevResearch.2.043077}
}

@article{Farrow2018,
author = {Pusuluk, Onur  and Farrow, Tristan  and Deliduman, Cemsinan  and Burnett, Keith  and Vedral, Vlatko },
title = {Proton tunnelling in hydrogen bonds and its implications in an induced-fit model of enzyme catalysis},
journal = {Proceedings of the Royal Society A: Mathematical, Physical and Engineering Sciences},
volume = {474},
number = {2218},
pages = {20180037},
year = {2018},
doi = {10.1098/rspa.2018.0037},
URL = {https://royalsocietypublishing.org/doi/abs/10.1098/rspa.2018.0037}
}

@article{Rabi1936,
  title = {On the Process of Space Quantization},
  author = {Rabi, I. I.},
  journal = {Phys. Rev.},
  volume = {49},
  issue = {4},
  pages = {324--328},
  numpages = {0},
  year = {1936},
  month = {Feb},
  publisher = {American Physical Society},
  doi = {10.1103/PhysRev.49.324},
  url = {https://link.aps.org/doi/10.1103/PhysRev.49.324}
}

@article{Rabi1937,
  title = {Space Quantization in a Gyrating Magnetic Field},
  author = {Rabi, I. I.},
  journal = {Phys. Rev.},
  volume = {51},
  issue = {8},
  pages = {652--654},
  numpages = {0},
  year = {1937},
  month = {Apr},
  publisher = {American Physical Society},
  doi = {10.1103/PhysRev.51.652},
  url = {https://link.aps.org/doi/10.1103/PhysRev.51.652}
}

@ARTICLE{Jaynes1963,
  author={Jaynes, E. T. and Cummings, F. W.},
  journal={Proceedings of the IEEE}, 
  title={Comparison of quantum and semiclassical radiation theories with application to the beam maser}, 
  year={1963},
  volume={51},
  number={1},
  pages={89-109},
  doi={10.1109/PROC.1963.1664},
  url={https://doi.org/10.1109/PROC.1963.1664}
}

@article{Tavis1968,
  title = {Exact Solution for an $N$-Molecule---Radiation-Field {H}amiltonian},
  author = {Tavis, Michael and Cummings, Frederick W.},
  journal = {Phys. Rev.},
  volume = {170},
  issue = {2},
  pages = {379--384},
  numpages = {0},
  year = {1968},
  month = {Jun},
  publisher = {American Physical Society},
  doi = {10.1103/PhysRev.170.379},
  url = {https://link.aps.org/doi/10.1103/PhysRev.170.379}
}

@article{Angelakis2007,
  title = {Photon-blockade-induced {M}ott transitions and {$XY$} spin models in coupled cavity arrays},
  author = {Angelakis, Dimitris G. and Santos, Marcelo Franca and Bose, Sougato},
  journal = {Phys. Rev. A},
  volume = {76},
  issue = {3},
  pages = {031805},
  numpages = {4},
  year = {2007},
  month = {Sep},
  publisher = {American Physical Society},
  doi = {10.1103/PhysRevA.76.031805},
  url = {https://link.aps.org/doi/10.1103/PhysRevA.76.031805}
}

@article{Smith2021,
  title = {Exact $k$-body representation of the {J}aynes--{C}ummings interaction in the dressed basis: {I}nsight into many-body phenomena with light},
  author = {Smith, Kevin C. and Bhattacharya, Aniruddha and Masiello, David J.},
  journal = {Phys. Rev. A},
  volume = {104},
  issue = {1},
  pages = {013707},
  numpages = {23},
  year = {2021},
  month = {Jul},
  publisher = {American Physical Society},
  doi = {10.1103/PhysRevA.104.013707},
  url = {https://link.aps.org/doi/10.1103/PhysRevA.104.013707}
}

@article{Dull2021,
  title = {Quality of Control in the {T}avis--{C}ummings--{H}ubbard Model},
  author = {Düll, R. and Kulagin, A. and Lee, L. and Ozhigov, Y. and Miao, H. and Zheng, K.},
  journal = {Computational Mathematics and Modeling},
  volume = {32},
  issue = {1},
  pages = {75-85},
  year = {2021},
  doi = {10.1007/s10598-021-09517-y},
  url = {https://doi.org/10.1007/s10598-021-09517-y}
}

@article{MiaoHuihui2024,
author = {Miao, Hui-hui},
title = {Investigating entropic dynamics of multiqubit cavity {QED} system},
journal = {Advanced Quantum Technologies},
volume = {7},
issue = {12},
pages = {2400246},
year = {2024},
doi = {10.1002/qute.202400246},
URL = {http://dx.doi.org/10.1002/qute.202400246},
}

@article{MiaoLi2025,
author = {Miao, Hui-hui and Li, Wanshun},
title = {Entanglement and quantum discord in the cavity {QED} models},
journal = {Heliyon},
volume = {11},
issue = {1},
pages = {e41194},
year = {2025},
doi = {10.1016/j.heliyon.2024.e41194},
URL = {https://doi.org/10.1016/j.heliyon.2024.e41194},
}

@article{You2025,
title = {Simulating and investigating various dynamic aspects of the {H}2{O}-related hydrogen bond model},
journal = {Chinese Journal of Physics},
volume = {97},
pages = {753-767},
year = {2025},
issn = {0577-9073},
doi = {https://doi.org/10.1016/j.cjph.2025.06.018},
url = {https://www.sciencedirect.com/science/article/pii/S0577907325002369},
author = {Jiangchuan You and Ran Chen and Wanshun Li and Hui-hui Miao and Yuri Igorevich Ozhigov}
}

@article{Lee1999,
  title = {Dark states of dressed {B}ose-{E}instein condensates},
  author = {Lee, E. S. and Geckeler, C. and Heurich, J. and Gupta, A. and Cheong, Kit-Iu and Secrest, S. and Meystre, P.},
  journal = {Phys. Rev. A},
  volume = {60},
  issue = {5},
  pages = {4006--4011},
  numpages = {0},
  year = {1999},
  month = {Nov},
  publisher = {American Physical Society},
  doi = {10.1103/PhysRevA.60.4006},
  url = {https://link.aps.org/doi/10.1103/PhysRevA.60.4006}
}

@article{Andre2002,
  title = {Coherent atom interactions mediated by dark-state polaritons},
  author = {André, A and Duan, L-M and Lukin, M D},
  journal = {Phys Rev Lett},
  volume = {88},
  issue = {24},
  pages = {243602},
  year = {2002},
  month = {Jun},
  doi = {10.1103/PhysRevLett.88.243602},
  url = {https://doi.org/10.1103/PhysRevLett.88.243602}
}

@article{Poltl2012,
author = {Pöltl, Christina and Emary, Clive and Brandes, Tobias},
year = {2012},
month = {09},
pages = {},
title = {Spin entangled two-particle dark state in quantum transport through coupled quantum dots},
volume = {87},
journal = {Physical Review B},
doi = {10.1103/PhysRevB.87.045416}
}

@article{Tanamoto2012,
author = {Tanamoto, Tetsufumi and Ono, Keiji and Nori, Franco},
year = {2012},
month = {02},
pages = {02BJ07},
title = {Steady-State Solution for Dark States Using a Three-Level System in Coupled Quantum Dots},
volume = {51},
journal = {Japanese Journal of Applied Physics},
doi = {10.7567/JJAP.51.02BJ07}
}

@article{Hansom2014,
  title = {Environment-assisted quantum control of a solid-state spin via coherent dark states},
  author = {Hansom, Jack and Schulte, Carsten H. H. and Le Gall, Claire and Matthiesen, Clemens and Clarke, Edmund and Hugues, Maxime and Taylor, Jacob M. and Atatüre, Mete},
  journal = {Nature Physics},
  volume = {10},
  issue = {10},
  pages = {725--730},
  numpages = {0},
  year = {2014},
  month = {Oct},
  publisher = {American Physical Society},
  doi = {10.1038/nphys3077},
  url = {https://doi.org/10.1038/nphys3077}
}

@article{Ozhigov2020,
  title = {Optical Selection of Dark States of Multilevel Atomic Ensembles},
  author = {Kulagin, A. V. and Ozhigov, Yu. I.},
  journal = {Computational Mathematics and Modeling},
  volume = {31},
  issue = {4},
  pages = {431-441},
  year = {2020},
  doi = {10.1007/s10598-021-09504-3},
  url = {https://doi.org/10.1007/s10598-021-09504-3}
}

@article{MiaoOzhigov2024,
author = {Miao, Hui-hui and Ozhigov, Yuri Igorevich},
title = {Distributed computing quantum unitary evolution},
journal = {Lobachevskii Journal of Mathematics},
volume = {45},
issue = {7},
pages = {3121-3129},
year = {2024},
doi = {10.1134/S1995080224603904},
URL = {http://dx.doi.org/10.1134/S1995080224603904},
}

@article{LiMiao2024,
author = {Li, Wanshun and Miao, Hui-hui and Ozhigov, Yuri Igorevich},
title = {Supercomputer model of finite-dimensional quantum electrodynamics applications},
journal = {Lobachevskii Journal of Mathematics},
volume = {45},
issue = {7},
pages = {3097-3106},
year = {2024},
doi = {10.1134/S1995080224603849},
URL = {http://dx.doi.org/10.1134/S1995080224603849},
}

@article{Wu2007,
  title = {Strong-Coupling Theory of Periodically Driven Two-Level Systems},
  author = {Wu, Ying and Yang, Xiaoxue},
  journal = {Phys. Rev. Lett.},
  volume = {98},
  issue = {1},
  pages = {013601},
  numpages = {4},
  year = {2007},
  month = {Jan},
  publisher = {American Physical Society},
  doi = {10.1103/PhysRevLett.98.013601},
  url = {https://link.aps.org/doi/10.1103/PhysRevLett.98.013601}
}

@article{Ozhigov2021,
  title = {About chemical modifications of finite dimensional {QED} models},
  author = {Vitaliy, Afanasyev and Zheng, Keli and Alexei, Kulagin and Huihui Miao and Yuri, Ozhigov and Wanshun, Lee and Nadezda, Victorova},
  journal = {Nonlinear Phenomena in Complex Systems},
  volume = {24},
  number = {3},
  pages = {230-241},
  year = {2021},
  url = {https://doi.org/10.33581/1561-4085-2021-24-3-230-241}
}

@article{Miao2023,
title = {Using a modified version of the {T}avis--{C}ummings--{H}ubbard model to simulate the formation of neutral hydrogen molecule},
journal = {Physica A: Statistical Mechanics and its Applications},
volume = {622},
pages = {128851},
year = {2023},
issn = {0378-4371},
doi = {https://doi.org/10.1016/j.physa.2023.128851},
url = {https://www.sciencedirect.com/science/article/pii/S0378437123004065},
author = {Hui-hui Miao and Yuri Igorevich Ozhigov}
}

@article{MiaoOzhigov2023,
author = {Miao, Hui-hui and Ozhigov, Yuri Igorevich},
title = {Comparing the effects of nuclear and electron spins on the formation of neutral hydrogen molecule},
journal = {Lobachevskii Journal of Mathematics},
volume = {44},
issue = {8},
pages = {3111-3120},
year = {2023},
doi = {10.1134/S1995080223080401},
URL = {https://doi.org/10.1134/S1995080223080401},
}

@article{Buck2000,
author = {Buck, Udo and Huisken, Friedrich},
title = {Infrared Spectroscopy of Size-Selected Water and Methanol Clusters},
journal = {Chemical Reviews},
volume = {100},
number = {11},
pages = {3863-3890},
year = {2000},
doi = {10.1021/cr990054v},
URL = {https://doi.org/10.1021/cr990054v},
}

@article{Xantheas1995,
    author = {Xantheas, Sotiris S.},
    title = {Ab initio studies of cyclic water clusters ({H}2{O})n, n=1–6. {III}. {C}omparison of density functional with {MP2} results},
    journal = {The Journal of Chemical Physics},
    volume = {102},
    number = {11},
    pages = {4505-4517},
    year = {1995},
    month = {03},
    issn = {0021-9606},
    doi = {10.1063/1.469499},
    url = {https://doi.org/10.1063/1.469499},
}

@article{Wang2011,
    author = {Wang, Yimin and Bowman, Joel M.},
    title = {Ab initio potential and dipole moment surfaces for water. {II.} {L}ocal-monomer calculations of the infrared spectra of water clusters},
    journal = {The Journal of Chemical Physics},
    volume = {134},
    number = {15},
    pages = {154510},
    year = {2011},
    month = {04},
    issn = {0021-9606},
    doi = {10.1063/1.3579995},
    url = {https://doi.org/10.1063/1.3579995},
}

@article{Ceriotti2016,
    author = {Ceriotti, M and Fang, W and Kusalik, PG and McKenzie, RH and Michaelides, A and Morales, MA and Markland, TE},
    title = { Nuclear Quantum Effects in Water and Aqueous Systems: Experiment, Theory, and Current Challenges.},
    journal = {Chem Rev.},
    volume = {116},
    pages = {7529-7550},
    year = {2016},
    doi = {10.1021/acs.chemrev.5b00674},
    url = {https://doi.org/10.1021/acs.chemrev.5b00674.},
}

@article{Mei1998,
author = {Mei, Hsiao S. and Tuckerman, Mark E. and Sagnella, Diane E. and Klein, Michael L.},
title = {Quantum Nuclear ab Initio Molecular Dynamics Study of Water Wires},
journal = {The Journal of Physical Chemistry B},
volume = {102},
number = {50},
pages = {10446-10458},
year = {1998},
doi = {10.1021/jp982623t},
URL = {https://doi.org/10.1021/jp982623t},
}

@article{Ceotto2017,
  title = {Semiclassical ``Divide-and-Conquer'' Method for Spectroscopic Calculations of High Dimensional Molecular Systems},
  author = {Ceotto, Michele and Di Liberto, Giovanni and Conte, Riccardo},
  journal = {Phys. Rev. Lett.},
  volume = {119},
  issue = {1},
  pages = {010401},
  numpages = {7},
  year = {2017},
  month = {Jul},
  publisher = {American Physical Society},
  doi = {10.1103/PhysRevLett.119.010401},
  url = {https://link.aps.org/doi/10.1103/PhysRevLett.119.010401}
}
	
\end{document}